\documentclass[conference,10pt,twoside,twocolumn]{IEEEtran}

\usepackage{cite}
\usepackage[T1]{fontenc}
\usepackage{graphicx}
\usepackage{amssymb}
\usepackage{amsmath}
\usepackage{amsthm}
\usepackage{subfigure}
\usepackage{booktabs} 
\usepackage{multirow}
\usepackage{microtype}
\usepackage{balance}
\usepackage{xcolor}
\usepackage{cite}
\usepackage{microtype}

\usepackage{amssymb}
\usepackage{amsfonts}
\usepackage{mathrsfs}
\usepackage{xspace}
\usepackage{bm}
\usepackage{upgreek}

\newcommand{\safemath}[2]{\newcommand{#1}{\ensuremath{#2}\xspace}}

\safemath{\bma}{\mathbf{a}}
\safemath{\bmb}{\mathbf{b}}
\safemath{\bmc}{\mathbf{c}}
\safemath{\bmd}{\mathbf{d}}
\safemath{\bme}{\mathbf{e}}
\safemath{\bmf}{\mathbf{f}}
\safemath{\bmg}{\mathbf{g}}
\safemath{\bmh}{\mathbf{h}}
\safemath{\bmi}{\mathbf{i}}
\safemath{\bmj}{\mathbf{j}}
\safemath{\bmk}{\mathbf{k}}
\safemath{\bml}{\mathbf{l}}
\safemath{\bmm}{\mathbf{m}}
\safemath{\bmn}{\mathbf{n}}
\safemath{\bmo}{\mathbf{o}}
\safemath{\bmp}{\mathbf{p}}
\safemath{\bmq}{\mathbf{q}}
\safemath{\bmr}{\mathbf{r}}
\safemath{\bms}{\mathbf{s}}
\safemath{\bmt}{\mathbf{t}}
\safemath{\bmu}{\mathbf{u}}
\safemath{\bmv}{\mathbf{v}}
\safemath{\bmw}{\mathbf{w}}
\safemath{\bmx}{\mathbf{x}}
\safemath{\bmy}{\mathbf{y}}
\safemath{\bmz}{\mathbf{z}}
\safemath{\bmzero}{\mathbf{0}}
\safemath{\bmone}{\mathbf{1}}

\bmdefine{\biad}{a}
\bmdefine{\bibd}{b}
\bmdefine{\bicd}{c}
\bmdefine{\bidd}{d}
\bmdefine{\bied}{e}
\bmdefine{\bifd}{f}
\bmdefine{\bigd}{g}
\bmdefine{\bihd}{h}
\bmdefine{\biid}{i}
\bmdefine{\bijd}{j}
\bmdefine{\bikd}{k}
\bmdefine{\bild}{l}
\bmdefine{\bimd}{m}
\bmdefine{\bind}{n}
\bmdefine{\biod}{o}
\bmdefine{\bipd}{p}
\bmdefine{\biqd}{q}
\bmdefine{\bird}{r}
\bmdefine{\bisd}{s}
\bmdefine{\bitd}{t}
\bmdefine{\biud}{u}
\bmdefine{\bivd}{v}
\bmdefine{\biwd}{w}
\bmdefine{\bixd}{x}
\bmdefine{\biyd}{y}
\bmdefine{\bizd}{z}

\bmdefine{\bixid}{\xi}
\bmdefine{\bilambdad}{\lambda}
\bmdefine{\bimud}{\mu}
\bmdefine{\bithetad}{\theta}
\bmdefine{\biphid}{\phi}
\bmdefine{\bideltad}{\delta}

\safemath{\bmia}{\biad}
\safemath{\bmib}{\bibd}
\safemath{\bmic}{\bicd}
\safemath{\bmid}{\bidd}
\safemath{\bmie}{\bied}
\safemath{\bmif}{\bifd}
\safemath{\bmig}{\bigd}
\safemath{\bmih}{\bihd}
\safemath{\bmii}{\biid}
\safemath{\bmij}{\bijd}
\safemath{\bmik}{\bikd}
\safemath{\bmil}{\bild}
\safemath{\bmim}{\bimd}
\safemath{\bmin}{\bind}
\safemath{\bmio}{\biod}
\safemath{\bmip}{\bipd}
\safemath{\bmiq}{\biqd}
\safemath{\bmir}{\bird}
\safemath{\bmis}{\bisd}
\safemath{\bmit}{\bitd}
\safemath{\bmiu}{\biud}
\safemath{\bmiv}{\bivd}
\safemath{\bmiw}{\biwd}
\safemath{\bmix}{\bixd}
\safemath{\bmiy}{\biyd}
\safemath{\bmiz}{\bizd}

\safemath{\bmxi}{\bixid}
\safemath{\bmlambda}{\bilambdad}
\safemath{\bmmu}{\bimud}
\safemath{\bmtheta}{\bithetad}
\safemath{\bmphi}{\biphid}
\safemath{\bmdelta}{\bideltad}

\safemath{\bA}{\mathbf{A}}
\safemath{\bB}{\mathbf{B}}
\safemath{\bC}{\mathbf{C}}
\safemath{\bD}{\mathbf{D}}
\safemath{\bE}{\mathbf{E}}
\safemath{\bF}{\mathbf{F}}
\safemath{\bG}{\mathbf{G}}
\safemath{\bH}{\mathbf{H}}
\safemath{\bI}{\mathbf{I}}
\safemath{\bJ}{\mathbf{J}}
\safemath{\bK}{\mathbf{K}}
\safemath{\bL}{\mathbf{L}}
\safemath{\bM}{\mathbf{M}}
\safemath{\bN}{\mathbf{N}}
\safemath{\bO}{\mathbf{O}}
\safemath{\bP}{\mathbf{P}}
\safemath{\bQ}{\mathbf{Q}}
\safemath{\bR}{\mathbf{R}}
\safemath{\bS}{\mathbf{S}}
\safemath{\bT}{\mathbf{T}}
\safemath{\bU}{\mathbf{U}}
\safemath{\bV}{\mathbf{V}}
\safemath{\bW}{\mathbf{W}}
\safemath{\bX}{\mathbf{X}}
\safemath{\bY}{\mathbf{Y}}
\safemath{\bZ}{\mathbf{Z}}

\safemath{\bZero}{\mathbf{0}}
\safemath{\bOne}{\mathbf{1}}
\safemath{\bDelta}{\mathbf{\Delta}}
\safemath{\bLambda}{\mathbf{\UpLambda}}
\safemath{\bPhi}{\mathbf{\Upphi}}
\safemath{\bSigma}{\mathbf{\Upsigma}}
\safemath{\bOmega}{\mathbf{\Upomega}}
\safemath{\bTheta}{\mathbf{\Uptheta}}

\bmdefine{\biAd}{A}
\bmdefine{\biBd}{B}
\bmdefine{\biCd}{C}
\bmdefine{\biDd}{D}
\bmdefine{\biEd}{E}
\bmdefine{\biFd}{F}
\bmdefine{\biGd}{G}
\bmdefine{\biHd}{H}
\bmdefine{\biId}{I}
\bmdefine{\biJd}{J}
\bmdefine{\biKd}{K}
\bmdefine{\biLd}{L}
\bmdefine{\biMd}{M}
\bmdefine{\biOd}{N}
\bmdefine{\biPd}{O}
\bmdefine{\biQd}{P}
\bmdefine{\biRd}{R}
\bmdefine{\biSd}{S}
\bmdefine{\biTd}{T}
\bmdefine{\biUd}{U}
\bmdefine{\biVd}{V}
\bmdefine{\biWd}{W}
\bmdefine{\biXd}{X}
\bmdefine{\biYd}{Y}
\bmdefine{\biZd}{Z}

\bmdefine{\biDelta}{\Delta}
\bmdefine{\biLambda}{\Lambda}
\bmdefine{\biPhi}{\Phi}
\bmdefine{\biSigma}{\Sigma}
\bmdefine{\biOmega}{\Omega}
\bmdefine{\biTheta}{\Theta}

\safemath{\bimA}{\biAd}
\safemath{\bimB}{\biBd}
\safemath{\bimC}{\biCd}
\safemath{\bimD}{\biDd}
\safemath{\bimE}{\biEd}
\safemath{\bimF}{\biFd}
\safemath{\bimG}{\biGd}
\safemath{\bimH}{\biHd}
\safemath{\bimI}{\biId}
\safemath{\bimJ}{\biJd}
\safemath{\bimK}{\biKd}
\safemath{\bimL}{\biLd}
\safemath{\bimM}{\biMd}
\safemath{\bimN}{\biNd}
\safemath{\bimO}{\biOd}
\safemath{\bimP}{\biPd}
\safemath{\bimQ}{\biQd}
\safemath{\bimR}{\biRd}
\safemath{\bimS}{\biSd}
\safemath{\bimT}{\biTd}
\safemath{\bimU}{\biUd}
\safemath{\bimV}{\biVd}
\safemath{\bimW}{\biWd}
\safemath{\bimX}{\biXd}
\safemath{\bimY}{\biYd}
\safemath{\bimZ}{\biZd}

\safemath{\bimDelta}{\biDelta}
\safemath{\bimLambda}{\biLambda}
\safemath{\bimPhi}{\biPhi}
\safemath{\bimSigma}{\biSigma}
\safemath{\bimOmega}{\biOmega}
\safemath{\bimTheta}{\biTheta}

\safemath{\setA}{\mathcal{A}}
\safemath{\setB}{\mathcal{B}}
\safemath{\setC}{\mathcal{C}}
\safemath{\setD}{\mathcal{D}}
\safemath{\setE}{\mathcal{E}}
\safemath{\setF}{\mathcal{F}}
\safemath{\setG}{\mathcal{G}}
\safemath{\setH}{\mathcal{H}}
\safemath{\setI}{\mathcal{I}}
\safemath{\setJ}{\mathcal{J}}
\safemath{\setK}{\mathcal{K}}
\safemath{\setL}{\mathcal{L}}
\safemath{\setM}{\mathcal{M}}
\safemath{\setN}{\mathcal{N}}
\safemath{\setO}{\mathcal{O}}
\safemath{\setP}{\mathcal{P}}
\safemath{\setQ}{\mathcal{Q}}
\safemath{\setR}{\mathcal{R}}
\safemath{\setS}{\mathcal{S}}
\safemath{\setT}{\mathcal{T}}
\safemath{\setU}{\mathcal{U}}
\safemath{\setV}{\mathcal{V}}
\safemath{\setW}{\mathcal{W}}
\safemath{\setX}{\mathcal{X}}
\safemath{\setY}{\mathcal{Y}}
\safemath{\setZ}{\mathcal{Z}}
\safemath{\emptySet}{\varnothing}

\safemath{\colA}{\mathscr{A}}
\safemath{\colB}{\mathscr{B}}
\safemath{\colC}{\mathscr{C}}
\safemath{\colD}{\mathscr{D}}
\safemath{\colE}{\mathscr{E}}
\safemath{\colF}{\mathscr{F}}
\safemath{\colG}{\mathscr{G}}
\safemath{\colH}{\mathscr{H}}
\safemath{\colI}{\mathscr{I}}
\safemath{\colJ}{\mathscr{J}}
\safemath{\colK}{\mathscr{K}}
\safemath{\colL}{\mathscr{L}}
\safemath{\colM}{\mathscr{M}}
\safemath{\colN}{\mathscr{N}}
\safemath{\colO}{\mathscr{O}}
\safemath{\colP}{\mathscr{P}}
\safemath{\colQ}{\mathscr{Q}}
\safemath{\colR}{\mathscr{R}}
\safemath{\colS}{\mathscr{S}}
\safemath{\colT}{\mathscr{T}}
\safemath{\colU}{\mathscr{U}}
\safemath{\colV}{\mathscr{V}}
\safemath{\colW}{\mathscr{W}}
\safemath{\colX}{\mathscr{X}}
\safemath{\colY}{\mathscr{Y}}
\safemath{\colZ}{\mathscr{Z}}

\safemath{\opA}{\mathbb{A}}
\safemath{\opB}{\mathbb{B}}
\safemath{\opC}{\mathbb{C}}
\safemath{\opD}{\mathbb{D}}
\safemath{\opE}{\mathbb{E}}
\safemath{\opF}{\mathbb{F}}
\safemath{\opG}{\mathbb{G}}
\safemath{\opH}{\mathbb{H}}
\safemath{\opI}{\mathbb{I}}
\safemath{\opJ}{\mathbb{J}}
\safemath{\opK}{\mathbb{K}}
\safemath{\opL}{\mathbb{L}}
\safemath{\opM}{\mathbb{M}}
\safemath{\opN}{\mathbb{N}}
\safemath{\opO}{\mathbb{O}}
\safemath{\opP}{\mathbb{P}}
\safemath{\opQ}{\mathbb{Q}}
\safemath{\opR}{\mathbb{R}}
\safemath{\opS}{\mathbb{S}}
\safemath{\opT}{\mathbb{T}}
\safemath{\opU}{\mathbb{U}}
\safemath{\opV}{\mathbb{V}}
\safemath{\opW}{\mathbb{W}}
\safemath{\opX}{\mathbb{X}}
\safemath{\opY}{\mathbb{Y}}
\safemath{\opZ}{\mathbb{Z}}
\safemath{\opZero}{\mathbb{O}}
\safemath{\identityop}{\opI}

\safemath{\veca}{\bma}
\safemath{\vecb}{\bmb}
\safemath{\vecc}{\bmc}
\safemath{\vecd}{\bmd}
\safemath{\vece}{\bme}
\safemath{\vecf}{\bmf}
\safemath{\vecg}{\bmg}
\safemath{\vech}{\bmh}
\safemath{\veci}{\bmi}
\safemath{\vecj}{\bmj}
\safemath{\veck}{\bmk}
\safemath{\vecl}{\bml}
\safemath{\vecm}{\bmm}
\safemath{\vecn}{\bmn}
\safemath{\veco}{\bmo}
\safemath{\vecp}{\bmp}
\safemath{\vecq}{\bmq}
\safemath{\vecr}{\bmr}
\safemath{\vecs}{\bms}
\safemath{\vect}{\bmt}
\safemath{\vecu}{\bmu}
\safemath{\vecv}{\bmv}
\safemath{\vecw}{\bmw}
\safemath{\vecx}{\bmx}
\safemath{\vecy}{\bmy}
\safemath{\vecz}{\bmz}

\safemath{\veczero}{\bmzero}
\safemath{\vecone}{\bmone}
\safemath{\vecxi}{\bmxi}
\safemath{\veclambda}{\bmlambda}
\safemath{\vecmu}{\bmmu}
\safemath{\vectheta}{\bmtheta}
\safemath{\vecphi}{\bmphi}
\safemath{\vecdelta}{\bmdelta}

\safemath{\matA}{\bA}
\safemath{\matB}{\bB}
\safemath{\matC}{\bC}
\safemath{\matD}{\bD}
\safemath{\matE}{\bE}
\safemath{\matF}{\bF}
\safemath{\matG}{\bG}
\safemath{\matH}{\bH}
\safemath{\matI}{\bI}
\safemath{\matJ}{\bJ}
\safemath{\matK}{\bK}
\safemath{\matL}{\bL}
\safemath{\matM}{\bM}
\safemath{\matN}{\bN}
\safemath{\matO}{\bO}
\safemath{\matP}{\bP}
\safemath{\matQ}{\bQ}
\safemath{\matR}{\bR}
\safemath{\matS}{\bS}
\safemath{\matT}{\bT}
\safemath{\matU}{\bU}
\safemath{\matV}{\bV}
\safemath{\matW}{\bW}
\safemath{\matX}{\bX}
\safemath{\matY}{\bY}
\safemath{\matZ}{\bZ}
\safemath{\matzero}{\bmzero}

\safemath{\matDelta}{\bDelta}
\safemath{\matLambda}{\bLambda}
\safemath{\matPhi}{\bPhi}
\safemath{\matSigma}{\bSigma}
\safemath{\matOmega}{\bOmega}
\safemath{\matTheta}{\bTheta}

\safemath{\matidentity}{\matI}
\safemath{\matone}{\matO}

\safemath{\rnda}{A}
\safemath{\rndb}{B}
\safemath{\rndc}{C}
\safemath{\rndd}{D}
\safemath{\rnde}{E}
\safemath{\rndf}{F}
\safemath{\rndg}{G}
\safemath{\rndh}{H}
\safemath{\rndi}{I}
\safemath{\rndj}{J}
\safemath{\rndk}{K}
\safemath{\rndl}{L}
\safemath{\rndm}{M}
\safemath{\rndn}{N}
\safemath{\rndo}{O}
\safemath{\rndp}{P}
\safemath{\rndq}{Q}
\safemath{\rndr}{R}
\safemath{\rnds}{S}
\safemath{\rndt}{T}
\safemath{\rndu}{U}
\safemath{\rndv}{V}
\safemath{\rndw}{W}
\safemath{\rndx}{X}
\safemath{\rndy}{Y}
\safemath{\rndz}{Z}

\safemath{\rveca}{\bimA}
\safemath{\rvecb}{\bimB}
\safemath{\rvecc}{\bimC}
\safemath{\rvecd}{\bimD}
\safemath{\rvece}{\bimE}
\safemath{\rvecf}{\bimF}
\safemath{\rvecg}{\bimG}
\safemath{\rvech}{\bimH}
\safemath{\rveci}{\bimI}
\safemath{\rvecj}{\bimJ}
\safemath{\rveck}{\bimK}
\safemath{\rvecl}{\bimL}
\safemath{\rvecm}{\bimM}
\safemath{\rvecn}{\bimN}
\safemath{\rveco}{\bomO}
\safemath{\rvecp}{\bimP}
\safemath{\rvecq}{\bimQ}
\safemath{\rvecr}{\bimR}
\safemath{\rvecs}{\bimS}
\safemath{\rvect}{\bimT}
\safemath{\rvecu}{\bimU}
\safemath{\rvecv}{\bimV}
\safemath{\rvecw}{\bimW}
\safemath{\rvecx}{\bimX}
\safemath{\rvecy}{\bimY}
\safemath{\rvecz}{\bimZ}

\safemath{\rvecxi}{\bmxi}
\safemath{\rveclambda}{\bmlambda}
\safemath{\rvecmu}{\bmmu}
\safemath{\rvectheta}{\bmtheta}
\safemath{\rvecphi}{\bmphi}

\safemath{\rmatA}{\bimA}
\safemath{\rmatB}{\bimB}
\safemath{\rmatC}{\bimC}
\safemath{\rmatD}{\bimD}
\safemath{\rmatE}{\bimE}
\safemath{\rmatF}{\bimF}
\safemath{\rmatG}{\bimG}
\safemath{\rmatH}{\bimH}
\safemath{\rmatI}{\bimI}
\safemath{\rmatJ}{\bimJ}
\safemath{\rmatK}{\bimK}
\safemath{\rmatL}{\bimL}
\safemath{\rmatM}{\bimM}
\safemath{\rmatN}{\bimN}
\safemath{\rmatO}{\bimO}
\safemath{\rmatP}{\bimP}
\safemath{\rmatQ}{\bimQ}
\safemath{\rmatR}{\bimR}
\safemath{\rmatS}{\bimS}
\safemath{\rmatT}{\bimT}
\safemath{\rmatU}{\bimU}
\safemath{\rmatV}{\bimV}
\safemath{\rmatW}{\bimW}
\safemath{\rmatX}{\bimX}
\safemath{\rmatY}{\bimY}
\safemath{\rmatZ}{\bimZ}

\safemath{\rmatDelta}{\bimDelta}
\safemath{\rmatLambda}{\bimLambda}
\safemath{\rmatPhi}{\bimPhi}
\safemath{\rmatSigma}{\bimSigma}
\safemath{\rmatOmega}{\bimOmega}
\safemath{\rmatTheta}{\bimTheta}

\usepackage{amssymb}
\usepackage{amsfonts}
\usepackage{mathrsfs}
\usepackage{xspace}
\usepackage{bm}
\usepackage{fancyref}
\usepackage{textcomp}

\usepackage{multirow}
\usepackage{stmaryrd}

\newenvironment{textbmatrix}{	\setlength{\arraycolsep}{2.5pt}%
								\big[\begin{matrix}}{\end{matrix}\big]%
								\raisebox{0.08ex}{\vphantom{M}}}

\def\be{\begin{equation}}
\def\ee{\end{equation}}
\def\een{\nonumber \end{equation}}
\def\mat{\begin{bmatrix}}
\def\emat{\end{bmatrix}}
\def\btm{\begin{textbmatrix}}
\def\etm{\end{textbmatrix}}

\def\ba#1\ea{\begin{align}#1\end{align}}
\def\bas#1\eas{\begin{align*}#1\end{align*}}
\def\bs#1\es{\begin{split}#1\end{split}}
\def\bg#1\eg{\begin{gather}#1\end{gather}}
\def\bml#1\eml{\begin{multline}#1\end{multline}}
\def\bi#1\ei{\begin{itemize}#1\end{itemize}}

\DeclareMathOperator{\sign}{sign}			
\DeclareMathOperator{\Prob}{\opP}			

\safemath{\dirac}{\delta}					
\safemath{\krond}{\dirac}					

\safemath{\upto}{\uparrow}
\safemath{\downto}{\downarrow}
\safemath{\iu}{j}							
\safemath{\ev}{\lambda}						
\safemath{\hilseqspace}{l^{2}}				
\newcommand{\banachfunspace}[1]{\setL^{#1}}	
\safemath{\hilfunspace}{\banachfunspace{2}}	

\safemath{\SNR}{\textit{SNR}} 				
\safemath{\PAR}{\textit{PAR}} 				
\safemath{\No}{N_0}							
\safemath{\Es}{E_s}							
\safemath{\Eb}{E_b}							
\safemath{\EbNo}{\frac{\Eb}{\No}}
\safemath{\EsNo}{\frac{\Es}{\No}}

\DeclareMathOperator{\CHop}{\ensuremath{\opH}} 
\safemath{\tvir}{\rndh_{\CHop}}				
\safemath{\tvtf}{\rndl_{\CHop}}				
\safemath{\spf}{\rnds_{\CHop}}				
\safemath{\bff}{H_{\CHop}}					

\safemath{\ircf}{r_{h}}						
\safemath{\tftvcf}{r_{s}}					
\safemath{\tfcf}{r_{l}}						
\safemath{\bfcf}{r_{H}}						

\safemath{\tcorr}{c_h}						
\safemath{\scf}{c_{s}}						
\safemath{\tfcorr}{c_{l}}					
\safemath{\fcorr}{c_{H}}						

\safemath{\mi}{I}							
\safemath{\capacity}{C}						

\safemath{\normal}{\mathcal{N}}			
\safemath{\jpg}{\mathcal{CN}}			
\safemath{\mchain}{\leftrightarrow}		

\safemath{\dB}{\,\mathrm{dB}}
\safemath{\dBm}{\,\mathrm{dBm}}
\safemath{\Hz}{\,\mathrm{Hz}}
\safemath{\kHz}{\,\mathrm{kHz}}
\safemath{\MHz}{\,\mathrm{MHz}}
\safemath{\GHz}{\,\mathrm{GHz}}
\safemath{\s}{\,\mathrm{s}}
\safemath{\ms}{\,\mathrm{ms}}
\safemath{\mus}{\,\mathrm{\text{\textmu}s}}
\safemath{\ns}{\,\mathrm{ns}}
\safemath{\ps}{\,\mathrm{ps}}
\safemath{\meter}{\,\mathrm{m}}
\safemath{\mm}{\,\mathrm{mm}}
\safemath{\cm}{\,\mathrm{cm}}
\safemath{\m}{\,\mathrm{m}}
\safemath{\W}{\,\mathrm{W}}
\safemath{\mW}{\, \mathrm{mW}}
\safemath{\J}{\,\mathrm{J}}
\safemath{\K}{\,\mathrm{K}}
\safemath{\bit}{\,\mathrm{bit}}
\safemath{\nat}{\,\mathrm{nat}}

\safemath{\define}{\triangleq}			

\safemath{\equivalent}{\sim}
\safemath{\distas}{\sim}					
\safemath{\sdiff}{\Delta}				

\safemath{\reals}{\mathbb{R}}
\safemath{\positivereals}{\reals_{+}}
\safemath{\integers}{\mathbb{Z}}
\safemath{\posint}{\integers_{+}}
\safemath{\naturals}{\mathbb{N}}
\safemath{\posnaturals}{\naturals_{+}}
\safemath{\complexset}{\mathbb{C}}
\safemath{\rationals}{\mathbb{Q}}

\newcommand*{\fancyrefapplabelprefix}{app}		
\newcommand*{\fancyrefthmlabelprefix}{thm}		
\newcommand*{\fancyreflemlabelprefix}{lem}		
\newcommand*{\fancyrefcorlabelprefix}{cor}		
\newcommand*{\fancyrefdeflabelprefix}{def}		
\newcommand*{\fancyrefproplabelprefix}{prop}		
\newcommand*{\fancyrefexmpllabelprefix}{exmpl}
\newcommand*{\fancyrefalglabelprefix}{alg}		
\newcommand*{\fancyreftbllabelprefix}{tbl}		

\frefformat{vario}{\fancyrefseclabelprefix}{Sec.~#1}
\frefformat{vario}{\fancyrefthmlabelprefix}{Thm.~#1}
\frefformat{vario}{\fancyreftbllabelprefix}{Tbl.~#1}
\frefformat{vario}{\fancyreflemlabelprefix}{Lem.~#1}
\frefformat{vario}{\fancyrefcorlabelprefix}{Cor.~#1}
\frefformat{vario}{\fancyrefdeflabelprefix}{Def.~#1}
\frefformat{vario}{\fancyreffiglabelprefix}{Fig.~#1}
\frefformat{vario}{\fancyrefapplabelprefix}{App.~#1}
\frefformat{vario}{\fancyrefeqlabelprefix}{(#1)}
\frefformat{vario}{\fancyrefproplabelprefix}{Prop.~#1}
\frefformat{vario}{\fancyrefexmpllabelprefix}{Ex.~#1}
\frefformat{vario}{\fancyrefalglabelprefix}{Alg.~#1}

\safemath{\dictab}{[\,\dicta\,\,\dictb\,]}

\safemath{\ysig}{\bmy}
\safemath{\ysighat}{\hat{\ysig}}
\safemath{\ysigdim}{M}
\safemath{\xsig}{\bmx}
\safemath{\xsigdim}{N}
\safemath{\nx}{n_x}
\safemath{\zsig}{\bmz}
\safemath{\zsigdim}{\ysigdim}
\safemath{\rsig}{\bmr}
\safemath{\Adict}{\bA}
\safemath{\Adicttilde}{\widetilde{\Adict}}
\safemath{\Adictdim}{\outputdim\times\xsigdim}
\safemath{\avec}{\bma}
\safemath{\avectilde}{\tilde{\avec}}
\safemath{\Bdict}{\bB}
\safemath{\Bdicttilde}{\widetilde{\Bdict}}
\safemath{\Cdict}{\bC}
\safemath{\cvec}{\bmc}
\safemath{\Ddict}{\bD}
\safemath{\Ddictdim}{\ysigdim\times\xsigdim}
\safemath{\dvec}{\bmd}
\safemath{\Ddicttilde}{\widetilde{\bD}}
\safemath{\Bonb}{\bB}
\safemath{\bvec}{\bmb}
\safemath{\Bonbdim}{\ysigdim\times\ysigdim}
\safemath{\noise}{\bmn}
\safemath{\noisedim}{\ysigim}
\safemath{\err}{\bme}
\safemath{\errdim}{\ysigdim}
\safemath{\errset}{\setE}
\safemath{\nerr}{n_e}
\safemath{\delop}{\bP_\errset}
\safemath{\delopc}{\bP_{{\errset}^c}}

\safemath{\cplxi}{\imath}
\safemath{\cplxj}{\jmath}

\safemath{\dict}{\matD}
\safemath{\inputdim}{N}		
\safemath{\outputdim}{M}		
\safemath{\sparsity}{S}	
\safemath{\inputdimA}{{N_a}}	
\safemath{\inputdimB}{{N_b}}	
\safemath{\elemA}{{n_a}}	
\safemath{\elemB}{{n_b}}	
\safemath{\resA}{\matR_a}	
\safemath{\resB}{\matR_b}	
\safemath{\subD}{\matS} 
\safemath{\subA}{\matS_a} 
\safemath{\subB}{\matS_b} 
\safemath{\dicta}{\matA} 	
\safemath{\dictb}{\matB} 	
\safemath{\hollowS}{H}
\safemath{\hollowA}{H_a}
\safemath{\hollowB}{H_b}
\safemath{\cross}{Z}
\safemath{\coh}{\mu_d}			
\safemath{\coha}{\mu_a}			
\safemath{\cohb}{\mu_b}			
\safemath{\mubs}{\nu}	
\safemath{\cohm}{\mu_m} 
\safemath{\dictset}{\setD}	
\safemath{\dictsetp}{\dictset(\coh,\coha,\cohb)}	
\safemath{\dictsetgen}{\dictset_\text{gen}}
\safemath{\dictsetgenp}{\dictsetgen(\coh)}
\safemath{\dictsetonb}{\dictset_\text{onb}}
\safemath{\dictsetonbp}{\dictsetonb(\coh)}

\safemath{\leftside}{U}
\safemath{\rightsideA}{R_a}
\safemath{\rightsideB}{R_b}

\safemath{\indexS}{\setI_S} 

\safemath{\na}{n_a}			
\safemath{\nb}{n_b}			
\safemath{\coeffa}{p_i}	
\safemath{\coeffb}{q_j}	
\safemath{\seta}{\setP}		
\safemath{\setb}{\setQ}     
\safemath{\setw}{\setW}	
\safemath{\setz}{\setZ}	
\safemath{\cola}{\veca}		
\safemath{\colb}{\vecb}		
\safemath{\cold}{\vecd}		
\safemath{\inputvec}{\vecx} 	
\safemath{\error}{\vece}	
\safemath{\noiseout}{\vecz} 	
\safemath{\inputvecel}{x}
\safemath{\inputveca}{\vecx_a}
\safemath{\inputvecb}{\vecx_b}
\safemath{\outputvec}{\vecy}	
\safemath{\lambdamin}{\lambda_{\mathrm{min}}}

\safemath{\elltwo}{\ell_2}
\safemath{\ellone}{\ell_1}
\safemath{\ellzero}{\ell_0}
\safemath{\ellinf}{\ell_\infty}
\safemath{\ellinftilde}{\ell_{\widetilde\infty}}
\safemath{\licard}{Z(\coh,\coha,\cohb)}
\safemath{\xsol}{\hat{x}}
\safemath{\xbord}{x_b}		
\safemath{\xstat}{x_s}		
\safemath{\xstatLone}{\tilde{x}_s}
\safemath{\order}{\mathcal{O}} 
\safemath{\scales}{\Theta} 
\safemath{\ones}{\mathbf{1}} 
\safemath{\zeroes}{\mathbf{0}} 
\safemath{\thlone}{\kappa(\coh,\cohb)} 
\safemath{\constoneA}{\delta} 
\safemath{\constoneB}{\epsilon} 
\safemath{\nlarge}{L}				   
\safemath{\sumlarge}{S_\nlarge}
\safemath{\maxlarger}{P_\nlarge}	   
\safemath{\Pzero}{\textrm{P0}}	
\safemath{\Pone}{\textrm{P1}}
\safemath{\vecfir}{\vecw}			 
\safemath{\vecsec}{\vecz}
\safemath{\elvecfir}{w}              
\safemath{\elvecsec}{z}				 
\safemath{\nlargefir}{n}
\safemath{\normout}{\gamma}
\safemath{\auxfun}{h}
\safemath{\supp}{\textrm{supp}}

\safemath{\indexa}{\ell}
\safemath{\indexb}{r}
\safemath{\indexc}{i}
\safemath{\indexd}{j}

\safemath{\project}{P}

\usepackage{framed}

\IEEEoverridecommandlockouts
\allowdisplaybreaks

\begin{document}

\title{Fixed-Throughput GRAND with FIFO Scheduling}

\author{\IEEEauthorblockN{Filippo Christen, Darja Nonaca, and Christoph Studer}\\[0.0cm]
\IEEEauthorblockA{\em Department of Information Technology and Electrical Engineering, ETH Zurich, Switzerland}
\textit{email: fchrist@student.ethz.ch, dnonaca@iis.ee.ethz.ch, and studer@ethz.ch}
\thanks{FC and DN contributed equally to this work. The work of DN and CS was supported in part by an ETH grant, and has received funding from the Swiss State Secretariat for Education, Research, and Innovation (SERI) under the SwissChips initiative. The authors thank J\'er\'emy Guichemerre and Chao Ji for valuable input and feedback.}
}

\maketitle

\begin{abstract}
Guessing random additive noise decoding (GRAND) is a code-agnostic decoding method that iteratively guesses the noise pattern affecting the received codeword. The number of noise sequences to test depends on the noise realization. Thus, GRAND exhibits random runtime which results in nondeterministic throughput. 
However, real-time systems must process the incoming data at a fixed rate, necessitating a fixed-throughput decoder in order to avoid losing data. 
We propose a first-in first-out (FIFO) scheduling architecture that enables a fixed throughput while improving the block error rate (BLER) compared to the common approach of imposing a maximum runtime constraint per received codeword. 
Moreover, we demonstrate that the \emph{average throughput} metric of GRAND-based hardware implementations typically provided in the literature can be misleading as one needs to operate at approximately one order of magnitude lower throughput to achieve the BLER of an unconstrained decoder.

\end{abstract}

\section{Introduction}

Guessing random additive noise decoding (GRAND)~\cite{KD19} is an emerging maximum-likelihood decoding method that iteratively guesses the corrupting noise sequences in descending order of their likelihood. 
The number of noise sequences to be tested depends on the instantaneous noise realization, which results in a nondeterministic (i.e., random) decoding throughput. However, hardware implementations for real-time applications must operate at a fixed throughput in order to sustain a fixed input data rate without losing data. 

The common approach to achieve constant throughput for GRAND-based hardware implementations is to impose a decoding time limit for each code block. This approach is called GRAND with abandonment (GRANDAB)~\cite{KD19}, which can be accomplished by imposing an upper limit on the number of noise sequences to be tested (assuming a constant number of tests is executed per time unit). The choice of this maximum decoding time limit results in a trade-off between achievable throughput and block error rate (BLER), as terminating GRAND early can result in additional decoding errors. 
To illustrate this trade-off, \fref{fig:intro_tp} shows the  \emph{average throughput} (black $\boldsymbol{\times}$) achieved at $1\%$ BLER by the hardware design D1 from~\cite{CJ23} of ordered reliability bits GRAND (ORBGRAND)~\cite{KD22}, assuming that no runtime constraints were imposed. 
By imposing a hard limit on the number of noise sequences to be tested with the same hardware design (red curve with $\blacktriangle$ markers) one can realize a tradeoff between \emph{constant throughput} and minimum $E_\text{b}/N_\text{0}$ (in dB) required to reach 1\% BLER. 
From \fref{fig:intro_tp}, we observe that a real-time system with constant throughput can achieve a 1\% BLER only at an $E_\text{b}/N_\text{0}$ of 0.57\,dB higher than the $E_\text{b}/N_\text{0}$ predicted by an unconstrained (i.e., without imposing any runtime limit) ORBGRAND implementation. Similarly, to achieve the same $E_\text{b}/N_\text{0}$ at 1\% BLER as that of unconstrained ORBGRAND, a real-time system would suffer a throughput loss of approximately $21\times$.

\subsection{Contributions}
\begin{figure}[tp]
  \centering
  \includegraphics[width=0.75\columnwidth]{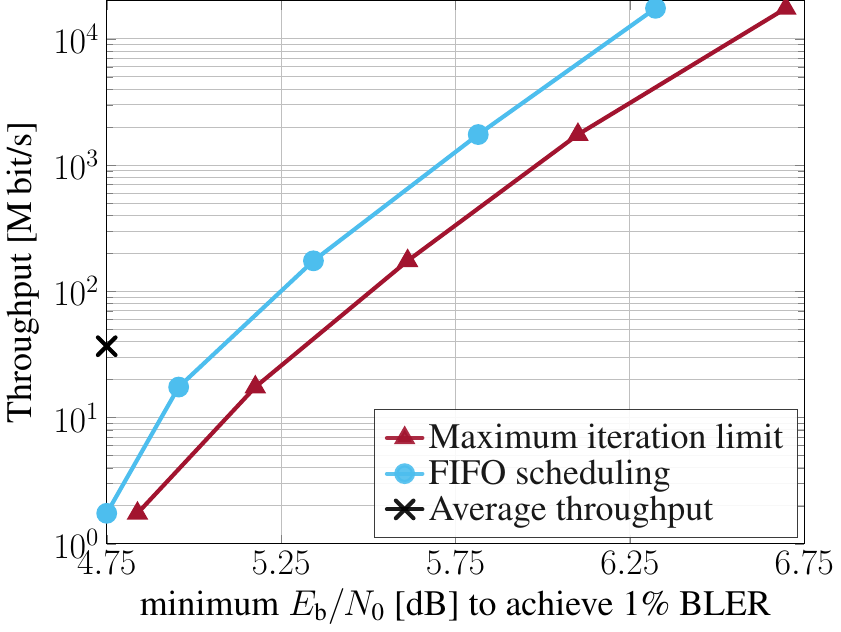}
  \caption{\emph{Average} throughput of ORBGRAND to achieve $1\%$ BLER versus maximum \emph{constant} throughput achieved by hardware design D1 from~\cite{CJ23} without FIFO scheduling (in red) and with FIFO scheduling (in blue).
  }
\label{fig:intro_tp}
\end{figure}

In order to achieve constant throughput as required by real-time systems while improving the BLER, we propose a FIFO-scheduling architecture that consists of an input first-in first-out (FIFO) buffer, one decoder or an array of parallel decoders, and an output re-order buffer (ROB). The input FIFO buffer stores received codewords and passes those codewords to available decoders.
The decoders considered in this paper implement ORBGRAND~\cite{KD22}.
The output ROB enables the decoder to store the recently decoded codewords, ready to be fetched in the correct order by the subsequent hardware block. 
In contrast to na\"ive approaches that impose a maximum iteration limit per codeword, our FIFO-scheduling architecture achieves (i) constant throughput and (ii) improved BLER by leveraging GRAND's random runtime which allows for higher runtime limits per codeword (see the blue curve with \scalebox{1.2}{$\bullet$} markers in \fref{fig:intro_tp}) at the cost of increased decoding latency. 
In order to demonstrate the efficacy of the proposed FIFO-scheduling architecture, we provide BLER, throughput, dynamic power, and latency results based on simulations with a random linear code and the ORBGRAND hardware design D1 from~\cite{CJ23}.

\subsection{Relevant Prior Work}

Several hardware implementations of ORBGRAND have been proposed recently~\cite{SA22,CJ23,abbas23,riaz21,blanc24}. Besides reporting the average throughput, these works also report the worst-case throughput---the true bottleneck for real-time applications. The work in~\cite{condo22} emphasizes the importance of considering the worst-case throughput when evaluating a decoder's performance, by improving the order of generated noise patterns to reduce the worst-case number of noise queries. %
In contrast to such ORBGRAND implementations, we propose a FIFO-scheduling architecture that leverages the random runtime of GRAND to improve error-rate performance. Our architecture achieves constant throughput while improving the BLER (see~the blue line in \fref{fig:intro_tp}) compared na\"ive solutions with runtime limits per code block. 
The work in~\cite{li21} proposes a multicore low-density parity check (LDPC) decoder architecture to increase the decoding throughput. In contrast, our architecture combines FIFO-scheduling with multicore processing in order to deal with a constant arrival rate of codewords while delivering a constant throughput (at a cost of increased latency). Our architecture can be used in combination with virtually any iterative decoder implementation that exhibits random runtime, such as (ORB)GRAND-based decoders and LDPC decoders.

\subsection{Notation}

Boldface lowercase and uppercase letters represent column vectors and matrices, respectively. 
For a vector $\bmx$, the $i$th entry is denoted by $x_i = [\bmx]_i$.
We denote conditional probabilities by $\Prob(\cdot|\cdot)$. 
For a $(n,k)$ linear block code, $n$ is the code length and $k$ the number of information bits; the code rate is $\varrho=k/n$ and the codebook is $\mathcal{C}$. A field of sequences of length $n$ is denoted $\mathbb{F}^n$. Addition in Galois field $\mathbb{F}_{2}$ is denoted by $\oplus$.

\section{Prerequisites}\label{sec:sysModel}

We consider forward error correction, where a block of $k$ information bits at the transmitter are mapped to an $n$-length codeword $\vecc \in \mathbb{F}_{2}^n$ selected from a codebook $\mathcal{C}$.
Each code bit~$c_i$, $i=1,\dots,n$, is transmitted over a memoryless additive white Gaussian noise (AWGN) channel with binary phase-shift keying (BPSK) symbols mapped according to $x_i=(-1)^{c_i}$.
The input-output relation of the AWGN channel is  $y_i = x_i + n_i$, where $y_i$ is the received signal and $n_i$ is real-valued zero-mean Gaussian noise with variance $\sigma^2$.
In what follows, we measure the signal-to-noise ratio in terms of $E_b/N_0=(\varrho\,\sigma^2)^{-1}$. 
We consider soft-input decoding, where the receiver first computes log-likelihood ratio (LLR) values for each coded bit as in~\cite{Cooper88}. 
Each block of LLR values $\lambda_i=\frac{2y_i}{\sigma^2}$, $i=1,\dots,n$, is then passed to, e.g., ORBGRAND~\cite{KD22} or our FIFO scheduling architecture, which generates estimates of the transmitted information bits.

ORBGRAND uses blocks of LLR values to iteratively guess the noise sequence $\vecz$ affecting the hard-demodulated received sequence $\hat{\vecy} = \frac{1}{2}(1-\sign(\vecy))$. More specifically, one generates noise sequences $\vecz$ in descending order of their likelihood (this is where the LLR values are being used) and iteratively tests these noise sequences by computing $\hat{\vecc} = \hat{\vecy} \oplus \vecz$ until a valid codeword is found, i.e., until $\hat{\vecc}\in \mathcal{C}$.

\section{Fixed-Throughput FIFO Scheduling}

Inspired by the FIFO scheduling architecture put forward in~\cite{CS09} in the context of sphere decoding for multi-antenna wireless communication systems, we now propose our FIFO scheduling architecture for GRAND-based decoders.

\subsection{FIFO Scheduling Architecture Overview}
\label{subsec:structure}

\begin{figure}[tp]
  \centering
  \includegraphics[width=0.98\columnwidth]{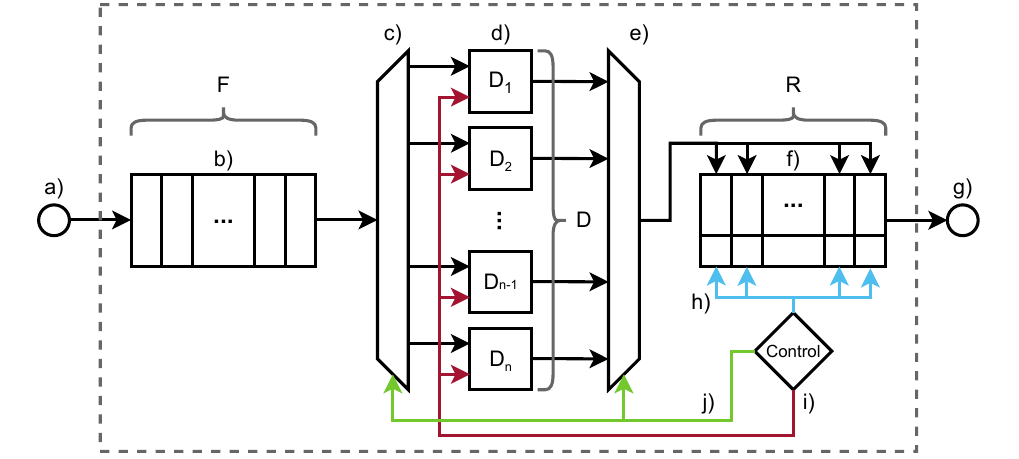}
  \vspace{-0.2cm}
  \caption{FIFO scheduling architecture overview: a) data source, b) input buffer, c) distribution unit, d) array of decoders, e) collection unit, f) ROB, g) data sink, h) booking mechanism, i) early termination, and j) distribution control.}
  \label{fig:structure}
\end{figure}
\fref{fig:structure} illustrates the general structure of our FIFO scheduling architecture. The operating principle is as follows: %
a)--b) Blocks of LLR values associated with each codeword entering the system are inserted into the input FIFO buffer of size $F$ (i.e., LLR values for up to $F$ received codewords can be stored). 
c)--d) A distribution unit passes one block of LLR values to an available decoder (chosen among $D$ decoder instances). 
e)--f) Once a codeword has been decoded (or the decoder was terminated early), a collection unit passes the codeword to the output re-order buffer (ROB), which can store up to $R$ decoded codewords. 
g) The decoded data is released from the ROB when it is requested by the subsequent hardware block.

This FIFO scheduling architecture allows a new decoding process to start as soon as a decoder becomes available, thereby reducing the decoders' idle time compared to an inflexible architecture in which a decoder must wait a fixed number of clock cycles before starting to decode a new codeword or terminating the decoding process early as soon as a new codeblock arrives at its input. Thus, FIFO scheduling exploits the decoder's random runtime, enabling some codewords to be processed longer than the average runtime, which reduces the BLER while maintaining constant throughput.

\subsection{ROB Booking Mechanism}\label{sec:booking}

Our FIFO scheduling architecture can be designed with an array of $D\geq 1$ decoders operating in parallel. Given the random decoding runtime of GRAND-based decoders, it is possible that the order of decoded codewords no longer corresponds to the same order as they arrived at the FIFO's input when multiple parallel decoders are employed.
In order to preserve the order of the received data, the data ordering within the input FIFO is mapped to the ROB through a process we call \textit{ROB booking mechanism}. Before leaving the input FIFO, each block of LLR values is assigned to the next available slot in the ROB in a sequential, queued manner. Each time a codeword is decoded, it is inserted into its reserved ROB slot, thereby retaining the original input data order at the output.

\begin{figure}[tp]
\centering
\subfigure[]{\includegraphics[scale =0.8]{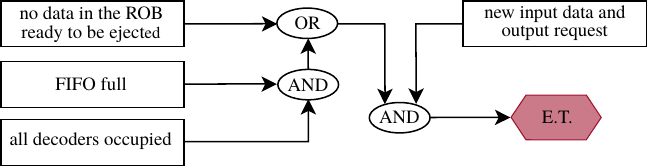}}\\
\hspace{-0.2cm}
\subfigure[]{\includegraphics[scale =0.8]{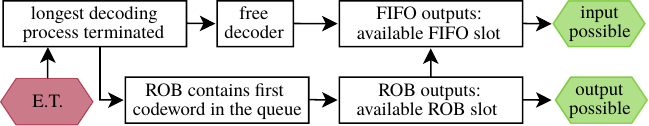}} 
\caption{(a) Conditions that trigger an early termination (E.~T.) and (b) control flow taking place when an early termination is triggered.}
\label{fig:ET}
\end{figure}

\subsection{Early Termination Mechanism}

In order to avoid data loss at the input, a free FIFO slot must be guaranteed whenever a new block of LLR values enters the input FIFO. Moreover, a processed codeword must be present within the ROB whenever it is expected at the output. We enforce both conditions by an \textit{early termination} mechanism, which terminates the longest-running decoding process early and forces the associated decoder to output its data. The conditions that lead to such an early termination and the control flow triggered by it are illustrated in \fref{fig:ET}.

\section{Results}\label{sec:results}

\subsection{Simulation Setup}
\label{subsubsec:Simulation}

We simulate the performance of different configurations of our FIFO scheduling architecture by processing $10^5$ codewords encoded with a random (256,234) linear block code.
The simulated configurations vary in FIFO size $F$, ROB size $R$ with $F=R\in \{1, 2, 4\}$, and $D\in \{1,2\}$ parallel decoders. We assume the ORBGRAND decoder design D1 from \cite{CJ23}. %
We define two simulation parameters: (i) the \textit{arrival interval}~$I$, measured in clock cycles per codeword, indicating the number of clock cycles after which a new codeword arrives at the decoder's input, and (ii) the \textit{data parallelism}~$P$, indicating the number of data vectors processed concurrently by the FIFO scheduling system. 
In order to limit the parameter space, we recall from \fref{sec:booking} that the input FIFO releases data only when the ROB has available slots. Therefore, the maximum data parallelism is $P_{\text{max}} = F+R$. We also ensure $R \geq D$ to keep all of the decoder cores busy.

\begin{figure}[tp]
  \centering
  \includegraphics[width=0.98\columnwidth]{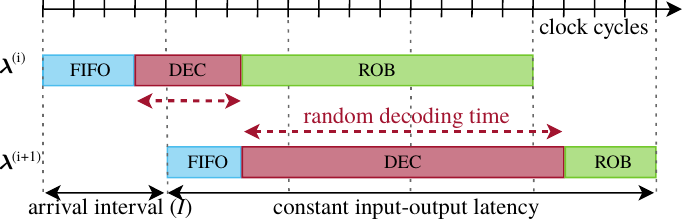}    
  \caption{FIFO scheduling with a single decoder, more than one FIFO/ROB slot ($F, R > 1$, $D=1$), $I=4$, and a constant input-output latency of $16$ clock cycles. Each row illustrates a hypothetical data flow through our architecture, showing a constant input-output latency for the $i$th and $(i+1)$th codeword~$\boldsymbol{\lambda}^\text{(i)}$ and $\boldsymbol{\lambda}^\text{(i+1)}$, respectively, despite the random decoding time.}
\label{fig:scheduling}
\end{figure}

In our system, a codeword is expelled from the ROB in the same clock cycle in which a new codeword arrives in the FIFO, both occurring every $I$ clock cycles. The FIFO scheduling system needs to accumulate some codewords before starting to output them in order to be able to fetch new data from the input FIFO once decoding has finished. 
Thus, a fixed input-output latency is enforced on the system during the initial $P\cdot I$ clock cycles of operation. After this initialization period, codewords are expelled from the ROB every $I$ clock cycles, ensuring constant throughput at the input and output interfaces. 
As shown in \fref{fig:scheduling}, this strategy results in a constant input-to-output latency, while internally, the decoding times can vary to accommodate ORBGRAND's random decoding time. 

\subsection{Performance Metrics}

We assess the efficacy of different architecture configurations by utilizing the algorithm and hardware specifications of the ORBGRAND decoder design D1 from~\cite{CJ23}. Concretely, we consider a clock frequency of $f=746$\,MHz, assume that the number of noise sequences tested by one decoder core per clock cycle is $\alpha=4$, and use a dynamic decoding power of $p_\text{dec} = 86.1\,\text{mW}$ per decoder. In our comparisons, we use the following performance metrics.

\begin{figure*}[tp]
\centering
\vspace{-0.25cm}
\hspace{-0.5cm}
\hfill
\subfigure[]{\includegraphics[width=0.32\linewidth]{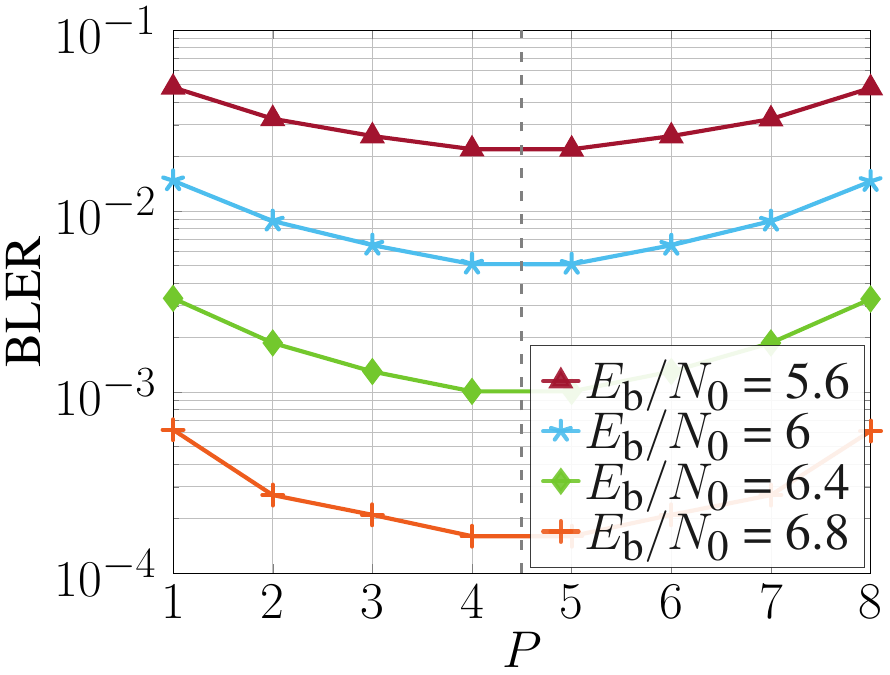} \label{fig:bler_diff_P}}
\hfill
\hspace{0.15cm}
  \subfigure[]{\includegraphics[width=0.31\linewidth]{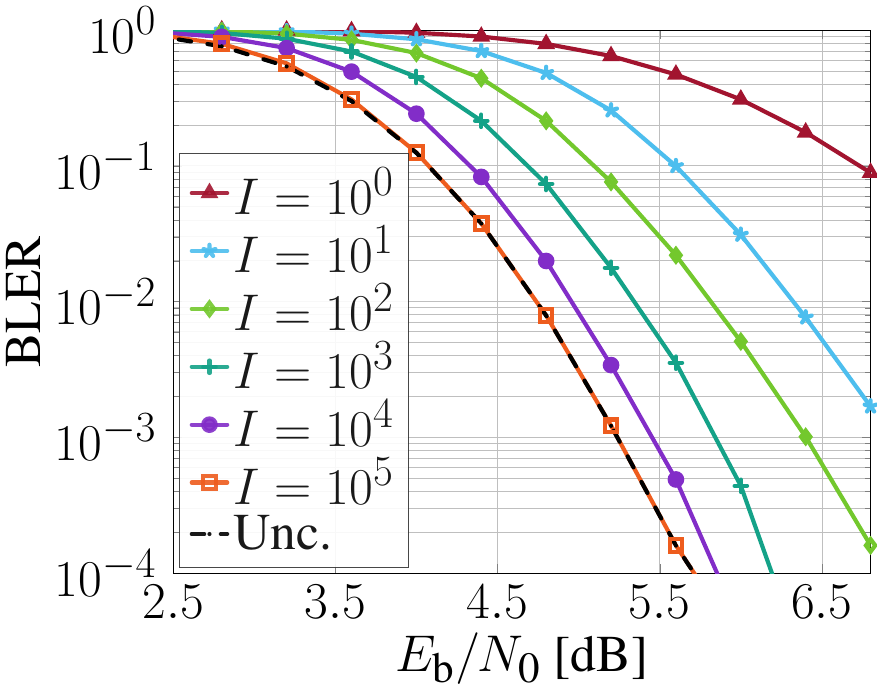} \label{fig:bler_diff_I}}
  \hfill
  \subfigure[]{\includegraphics[width=0.31\linewidth]{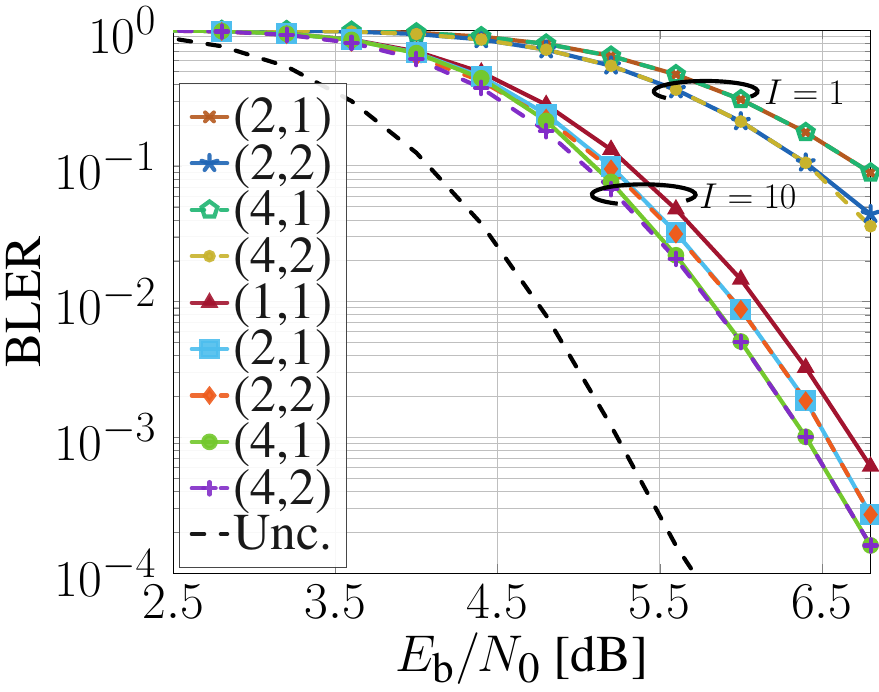}\label{fig:bler_diff_configs}}
\hfill
\vspace{-0.2cm}
\caption{Influence of data parallelism $P$ on BLER in a ($F=R=4$, $D=1$) FIFO scheduling architecture, $I = 10$ clock cycles (a). BLER at different arrival intervals $I$ in a (4,1) configuration and unconstrained (Unc.) ORBGRAND (b). BLER for different FIFO scheduling configurations ($F=R=P$, $D$) (c).}\label{fig:figs_bler}
\vspace{-0.5cm}
\end{figure*}

\begin{figure*}[tp]
\centering
	  \subfigure[]{\includegraphics[width=0.31\linewidth]{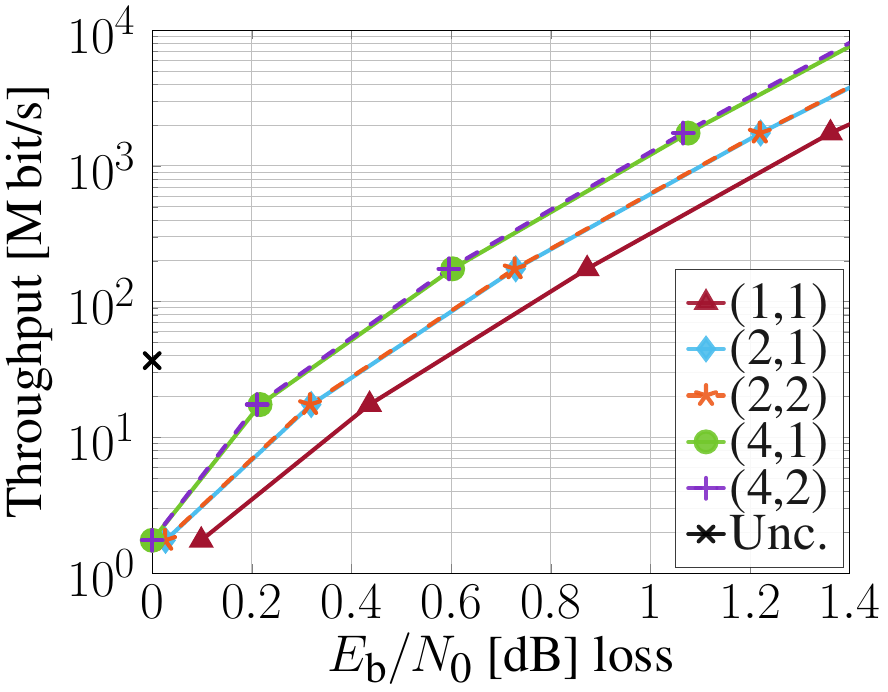} \label{fig:tp}}
\hfill
	\subfigure[]{\includegraphics[width=0.32\linewidth]{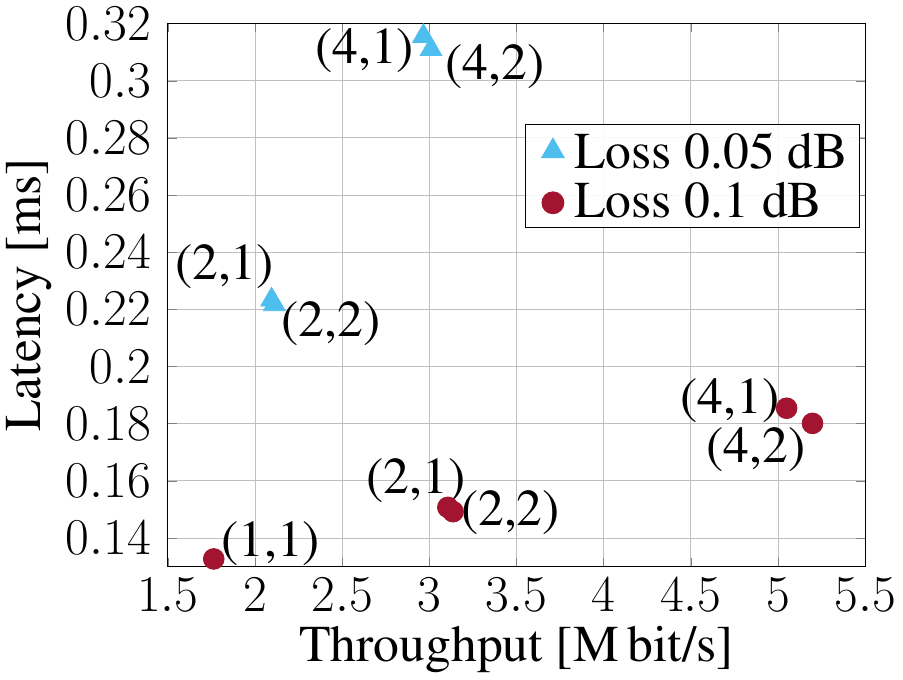}\label{fig:latency}}
\hfill
	\subfigure[]{\includegraphics[width=0.3\textwidth]{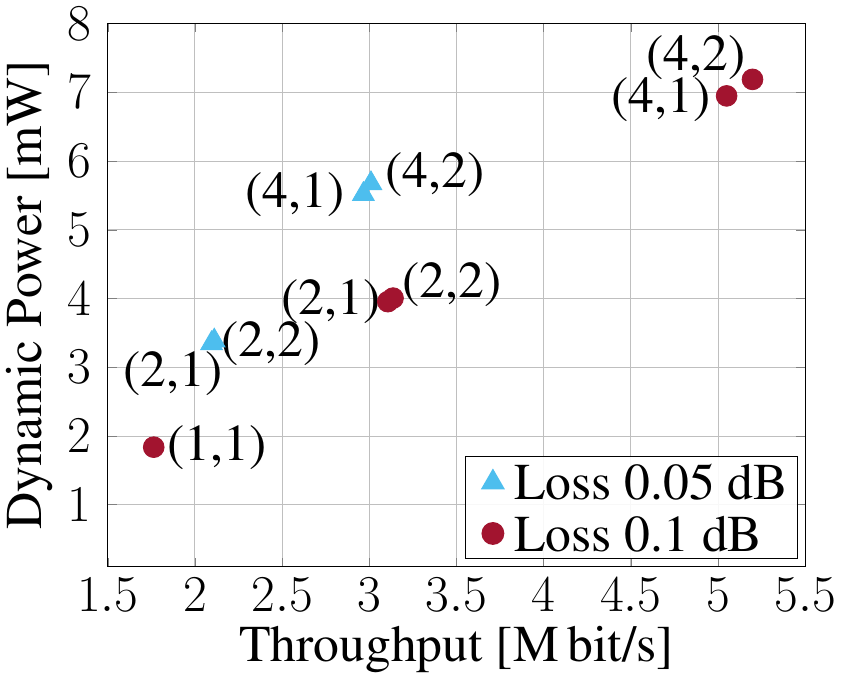}\label{fig:power}}
\vspace{-0.2cm}  
\caption{Maximum constant throughput achieved by different FIFO scheduling configurations ($F=R=P$, $D$) versus $E_\text{b}/N_\text{0}$ [dB] loss from 1\%\,BLER operating point achieved by an unconstrained ORBGRAND decoder (a). Latency (b) and dynamic power (c) versus throughput achieved by different FIFO scheduling configurations for 0.1 dB and 0.05 dB loss from 1\%\,BLER operating point achieved by an unconstrained ORBGRAND decoder.}
\end{figure*}

\subsubsection{Error-Rate Performance}
The BLER achieved by each configuration of our FIFO scheduling architecture is computed via $10^5$ Monte--Carlo trials per $E_\text{b}/N_\text{0}$ value.
The 1\%\,BLER operating point
is defined as the minimum $E_\text{b}/N_\text{0}$ (in decibels) required to achieve $1\%$ BLER.\footnote{Note that  $1\%$ BLER is conservative considering that the 3GPP 5G NR standard deems $2\%$ BLER acceptable~\cite[Table 8.1.1-1]{3gpp38133}.}
The $E_\text{b}/N_\text{0}$ loss is defined as the difference (in decibels) between the 1\%\,BLER operating point achieved by ORBGRAND without imposing a maximum iteration limit and the 1\%\,BLER operating point achieved by our FIFO scheduling architecture with a constant throughput.

\subsubsection{Throughput}
The throughput $\theta$ of our FIFO scheduling architecture is computed as $\theta = {kf}/{I}$\,bit/s.
The average throughput $\theta_{\text{avg}}$ refers to the throughput of an ORBGRAND decoder without runtime constraints and is computed as $\theta_\text{avg} = {\alpha k f}/{\beta}$\,bit/s, where $\beta$ is the average number of required iterations to decode a data vector at a given $E_\text{b}/N_\text{0}$. 

\subsubsection{Latency}
The constant input-output latency $\ell$ is an integer multiple of the arrival interval $I$, given by $\ell=(PI)/f$ seconds. 

\subsubsection{Dynamic Power}
To estimate the decoding power consumption $p$, we count the number of clock cycles $\delta_\text{act}(\mathrm{D_i})$ during which each decoder $\mathrm{D_i}$, $i=1,\ldots,D$ is actively decoding. We then compute the total number of clock cycles $\delta_\text{act,tot}=\sum_{i=1}^{D} \delta_\text{act}(\mathrm{D_i})$ during which the decoders are busy. 
We define the decoder's activity factor as $\eta_\text{act} = \delta_\text{act,tot}/(P I + I (10^5 - 1))$,  where the denominator is the number of clock cycles necessary to process all of the inputs. 
Finally, we estimate the dynamic decoding power as $p_\text{dyn} = \eta_\text{act} p_\text{dec}$ by ignoring control and FIFO/ROB power consumption.
\subsection{Results and Discussion}

We now discuss the design trade-offs in terms of BLER, throughput, latency, and power of different configurations of FIFO scheduling architecture. 

\fref{fig:bler_diff_P} shows the BLER for different data parallelisms $P$ and suggests the existence of an optimal $P_\text{opt}$ that minimizes the BLER for a given configuration of the FIFO scheduling architecture. Low $P$ causes frequent early terminations due to the ROB lacking the required processed codewords when requested;  high $P$ leads to input FIFO overflows, which also triggers early terminations. Thus, $P_\text{opt}$ must be chosen carefully.

\fref{fig:bler_diff_I} shows that increasing the arrival interval $I$ improves the BLER as it allows more decoding time per codeword. However, the BLER improvements diminish with larger $I$, converging to the BLER achieved by unconstrained ORBGRAND  at approximately $I = 10^5$. Thus, to reach the same BLER achieved by an unconstrained ORBGRAND design, a real-time system using a D1 instance from~\cite{CJ23} would need to tolerate latencies on the order of microseconds, which is impractical.

\fref{fig:bler_diff_configs} compares the BLER for different FIFO scheduling architecture configurations with fixed arrival intervals $I = 1$ and $I = 10$. 
For $I = 1$, adding a second decoder improves the BLER up to $0.3$\,dB as it lowers the occurrence of a full input FIFO. For $I = 10$, increasing buffer sizes yields up to $0.2$\,dB improvement, while a second decoder adds only minor gains as larger FIFO/ROB buffer sizes ensure sufficient processing time for a single decoder. 
Therefore, when the arrival interval is higher than the average number of iterations needed per codeword, it is more beneficial to increase the FIFO/ROB buffer sizes rather than adding decoder instances. 

\fref{fig:tp} shows a constant throughput achievable by different FIFO scheduling architectures ($F=R=P$, $D$) and compares them to the ORBGRAND's average throughput at the 1\% BLER operating point (black $\boldsymbol{\times}$). This comparison is in terms of the $E_\text{b}/N_\text{0}$ loss, which is the difference in decibels from the $\boldsymbol{\times}$ marker and the 1\% BLER operating point achievable by a given FIFO scheduling architecture. 
The simplest configuration (1,1), which is equivalent to the na\"ive maximum runtime limit per codeword, incurs a $0.6$\,dB loss; larger buffer sizes in the (4,1) configuration (reported also in \fref{fig:intro_tp}), reduce the loss to $0.35$\,dB. 
Moreover, to reach $0$\,dB loss with a (4,1) configuration, a real-time system would need to operate at a throughput that is $21\times$ lower than the average throughput of an unconstrained~decoder.

\fref{fig:latency} highlights the cost to pay in terms of input-output latency to achieve a constant target throughput with a small $E_\text{b}/N_\text{0}$ loss.
\fref{fig:power} shows the dynamic power of one or several decoder(s). We see that at higher throughputs, both the decoding latency and power consumption increase, as both quantities are proportional to the number of decoding iterations.

\section{Conclusions}

GRAND-based decoder implementations typically exhibit random runtime, but real-world communication systems must provide constant throughput and latency to avoid data loss.
In order to address this issue, we have proposed a novel FIFO scheduling architecture that guarantees constant throughput and latency, while outperforming a naïve approach that imposes a maximum decoding time per codeword. While our architecture increases latency and hardware overhead, our results reveal that increasing the input/output buffer sizes improves BLER performance for a wide throughput range, whereas increasing the number of parallel decoders instances only provides notable BLER gains at very high throughput. 
Furthermore, we have shown that achieving 1\% BLER under a constant-throughput requirement results in approximately one order of magnitude lower throughput compared to ORBGRAND’s average throughput, which is typically reported in the literature.

\bibliographystyle{IEEEtran}
\bibliography{bib/VIPabbrv,bib/confs-jrnls,bib/publishers,bib/VIP_190331, bib/24iscas_bib}

\end{document}